\documentclass[sigconf]{acmart}
\usepackage{enumitem}

\setlist[itemize]{leftmargin=*,nosep}

\AtBeginDocument{%
  }

\copyrightyear{2026}
\acmYear{2026}
\setcopyright{cc}
\setcctype{by}
\acmConference[RecSys '26]{20th ACM Conference on Recommender Systems}{September 27-October 02, 2026}{Minneapolis, MN, USA}
\acmBooktitle{20th ACM Conference on Recommender Systems (RecSys '26), September 27-October 02, 2026, Minneapolis, MN, USA}
\acmDOI{10.1145/3773078.3831930}
\acmISBN{979-8-4007-2284-4/2026/09}

\begin{document}

\title{RECAP: Feedback-Driven Streaming Semantic User Profiles for Short-Video Recommendation}

\author{Ziyi Zhao}
\affiliation{%
  \institution{University of Science and Technology of China}
  \city{Hefei}
  \state{Anhui}
  \country{China}
}
\email{re2477036@mail.ustc.edu.cn}


\author{Xiaoyou Zhou}
\email{zhouxiaoyou@kuaishou.com}
\author{Xiao Lv}
\authornote{Corresponding author.}
\email{lvxiao03@kuaishou.com}
\affiliation{%
  \institution{Kuaishou Technology}
  \city{Beijing}
  \country{China}
}


\author{Yangyang Li}
\affiliation{%
  \institution{China Academy of Cyber}
  \city{Beijing}
  \country{China}
}
\email{liyangyang@live.com}


\author{Chubo He}
\email{hechubo03@kuaishou.com}
\author{Zhao Liu}
\email{liuzhao09@kuaishou.com}
\affiliation{%
  \institution{Kuaishou Technology}
  \city{Beijing}
  \country{China}
}


\author{Jiayao Shen}
\email{shenjiayao@kuaishou.com}
\author{Yuqi Liu}
\email{liuyuqi10@kuaishou.com}
\affiliation{%
  \institution{Kuaishou Technology}
  \city{Beijing}
  \country{China}
}


\author{He Li}
\email{lihe12@kuaishou.com}
\author{Chengyi Zhang}
\email{zhangchengyi03@kuaishou.com}
\affiliation{%
  \institution{Kuaishou Technology}
  \city{Beijing}
  \country{China}
}


\author{Jian Liang}
\email{liangjian03@kuaishou.com}
\author{Ming Li}
\email{liming03@kuaishou.com}
\affiliation{%
  \institution{Kuaishou Technology}
  \city{Beijing}
  \country{China}
}


\author{Chongming Gao}
\authornotemark[1]
\email{chongming.gao@gmail.com}
\author{Fuli Feng}
\email{fulifeng93@gmail.com}
\affiliation{%
  \institution{University of Science and Technology of China}
  \city{Hefei}
  \state{Anhui}
  \country{China}
}


\author{Ruiming Tang}
\authornotemark[1]
\email{tangruiming@kuaishou.com}
\author{Han Li}
\email{lihan08@kuaishou.com}
\affiliation{%
  \institution{Kuaishou Technology}
  \city{Beijing}
  \country{China}
}

\renewcommand{\shortauthors}{Zhao et al.}

\begin{abstract}
Language-based user profiles convert long behavioral histories into explicit semantic representations for recommendation.
However, most profile generators are optimized in an open loop: they may summarize past behavior fluently, but are not directly trained to improve future recommendation.
We study this problem in real-world short-video recommendation, where user behaviors continuously arrive as streams and profiles must be incrementally updated under limited capacity.
This requires maintaining a consistent bounded profile state and constructing profile-targeted semantic feedback from industrial implicit behavior logs.
We propose RECAP, an offline closed-loop framework for optimizing streaming structured semantic profiles with historical recommendation feedback.
RECAP maintains each profile as a bounded structured memory by combining LLM-based semantic updates with deterministic lifecycle and capacity control.
RECAP constructs profile-targeted semantic feedback by filtering label-consistent behavior pairs with an LLM judge and training a dual-tower evaluator whose matching score serves as a GRPO reward.
Experiments on Kuaishou short-video data show that RECAP improves uAUC by 0.0084 and Recall@2000 by about 4.9\% over the base generator.
Further analyses confirm the benefits of feedback construction and policy optimization, and show more grounded refinement and user-level abstraction in profile updates.
A seven-day online A/B test further shows a statistically significant 0.139\% improvement in average application usage time per user.
\end{abstract}



\maketitle

\section{Introduction}

Traditional recommender systems rely on user behavior signals, ID embeddings, and interaction statistics to model user interests~\cite{hu2008collaborative,he2017neural,kang2018self,sun2019bert4rec}. 
While effective for behavioral co-occurrence, these representations often lack explicit descriptions of the semantic content behind user behaviors, such as topics users care about and interests connecting their interactions~\cite{zhang2020explainable}. 
Recent studies have explored using large language models (LLMs) to enhance recommender systems with semantic understanding and natural-language reasoning~\cite{zhao2023recommender,lin2023how}. 
Among them, language-based user profiles describe interests from user histories, providing interpretable features that enrich user representations and support personalization and cross-scenario transfer~\cite{xi2023kar,gao2024end,yi2025recgpttechnicalreport}.

When applying language-based profiles to industrial recommender systems, one-shot generation is often insufficient. 
In large-scale short-video recommendation, users may accumulate thousands of historical interactions, while new behaviors are continuously logged as users interact with the platform, causing their interests to evolve~\cite{wang2019sequential,zheng2024harnessing,zhu2024liber,xia2025hierarchical}. 
This naturally calls for streaming profile generation: behaviors are split into chronological chunks, and the LLM updates the existing profile after each new chunk~\cite{kim2025lost,zheng2024harnessing,zhu2024liber}. 
To support cross-round inheritance and localized revision, the profile should be maintained as a structured state rather than regenerated as a disposable free-text summary.
This streaming formulation raises a further question: how can such structured profiles be optimized from historical implicit feedback in industrial systems?

Recent studies have explored two relevant directions for language-based user profiling: feedback-driven profile optimization and long-history semantic modeling.
LangPTune and LettinGo optimize natural-language profiles with downstream recommendation signals~\cite{gao2024end,wang2025lettingo}, while LIBER and HiT-LBM improve long-history modeling by segmenting user behaviors and constructing or selecting high-quality interest representations~\cite{zhu2024liber,xia2025hierarchical}.
These studies demonstrate the value of recommendation-aligned profile generation and long-history semantic modeling.
However, they mainly focus on constructing profile texts, summaries, or interest representations, rather than closing the loop between raw implicit feedback and the policy that incrementally updates a bounded structured profile state.
How to optimize such a stateful profile updater in industrial short-video recommendation therefore remains underexplored.

Optimizing such a stateful profile updater faces two key challenges.
(1) \textit{Streaming profile-state maintenance}: the profile must preserve a reliable structured state over long-term, multi-round updates.
As user interests evolve under bounded capacity, the system must decide which interests to retain, update, merge, or remove while maintaining consistency.
(2) \textit{Profile-targeted semantic feedback construction}: optimizing the profile updater requires feedback that can assess the quality of generated profiles.
A straightforward choice is to reuse scores from a frozen industrial recommender, but its ranking scores are entangled with many features beyond profile semantics, which may dilute the signal for profile-item semantic alignment.
Moreover, logged watches and skips are implicit behavior labels, whose semantic meaning can be affected by content form, exposure context, or user inertia.
Thus, these raw signals do not directly correspond to profile-item semantic alignment.

To address these challenges, we propose RECAP, an offline closed-loop framework that recaps evolving user interests into bounded structured semantic profiles and optimizes profile updates with recommendation-aligned feedback.
(1) For streaming profile-state maintenance, RECAP combines an LLM semantic updater for interest recognition, topic merging, and description refinement with a deterministic state machine for strength decay, lifecycle transitions, and capacity control.
(2) For profile-targeted semantic feedback construction, RECAP first uses an LLM pairwise judge to identify behavior pairs whose implicit labels are consistent with semantic preferences, producing higher-confidence feedback from logged interactions.
(3) It then turns this feedback into a profile-targeted reward by training a dual-tower semantic evaluator over rendered profile text and video captions, using its profile-item matching scores as a semantic proxy reward for GRPO-based optimization of the profile updater.

Our contributions are summarized as follows.

\begin{itemize}
    \item We identify and study a practical problem in industrial recommender systems: optimizing streaming structured semantic profiles with downstream recommendation feedback.
    This problem extends profile generation from one-shot summarization to long-term structured updates, involving two key challenges: streaming profile-state maintenance and profile-targeted semantic feedback construction.

    \item We propose RECAP, an offline closed-loop framework that combines LLM-based semantic updates with a deterministic state machine, label-consistency-based feedback construction from logged implicit behaviors, and a dual-tower semantic evaluator whose scores guide GRPO-based policy optimization.

    \item We evaluate RECAP on real-world short-video recommendation data from Kuaishou.
    Offline experiments show improvements in uAUC and Recall, while further analyses confirm the benefits of feedback construction, profile capacity, and policy optimization.
    A seven-day online A/B test further confirms the effectiveness of RECAP in production.
\end{itemize}

\section{Related Work}

\subsection{Language-based User Profiles for Recommendation}

Language-based user profiling for recommendation has recently evolved along two directions: long-history profile construction and feedback-driven profile optimization.
For long-history construction, LLM-TRSR compresses segmented text-rich behavior sequences with hierarchical or recurrent LLM summarization for sequential recommendation~\cite{zheng2024harnessing}.
PersonaX selects representative sub-sequences and caches multiple personas to support recommendation agents with lower online profiling overhead~\cite{shi2025personax}.
LIBER partitions lifelong behaviors into short segments, summarizes intra-segment interests and inter-segment shifts, and fuses the resulting textual knowledge into recommendation backbones~\cite{zhu2024liber}.
HiT-LBM further improves cascaded chunk-level interest extraction by using process rating models and hierarchical tree search to select high-quality interest paths~\cite{xia2025hierarchical}.
These methods improve long-history compression and interest representation, but they mainly construct profile or interest representations as outputs of a generation pipeline.

For feedback-driven optimization, LangPTune trains an LLM profile encoder toward downstream ranking performance with a recommender decoder~\cite{gao2024end}.
LettinGo explores multiple free-form profiles and uses a downstream LLM predictor to construct preference pairs for profile alignment~\cite{wang2025lettingo}.
Unlike these methods, which optimize or select textual profile representations, RECAP treats the profile as a bounded structured interest state.
It focuses on how such a state should be maintained under streaming updates and optimized with high-confidence feedback from industrial implicit interactions.

\subsection{Long-term Memory for LLM-based User Modeling}

Long-term memory has been widely explored to help LLM systems preserve information beyond a single context window. 
Generative Agents and MemoryBank store, retrieve, and update experiences or user memories over long interactions~\cite{park2023generative,zhong2024memorybank}, while recent systems such as A-Mem and MemoryOS further organize memories with structured notes or modular memory operations~\cite{xu2025amem,kang2025memoryos}. 
These studies motivate treating user-related information as an explicit memory object that can be maintained over time. 
RECAP instantiates this idea in recommendation by maintaining semantic profiles as a bounded memory of structured interest entries over streaming user behaviors.

\begin{figure*}[t]
    \centering 
    \includegraphics[width=\linewidth]{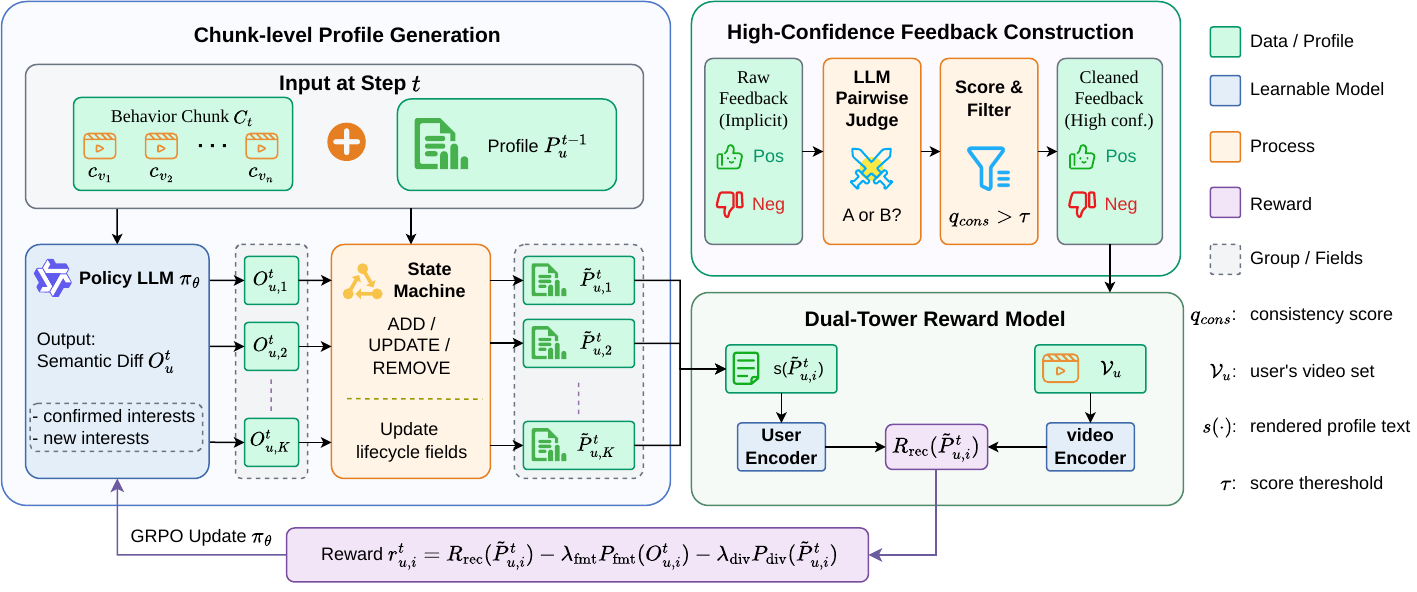}
    \Description{An overview of the RECAP framework for streaming structured semantic profile generation and feedback-driven optimization.}
    \caption{Overview of the RECAP framework.}
    \label{fig:overview}
\end{figure*}

\section{Preliminaries}

\subsection{Video Caption Generation}
\label{sec:video_caption}

To provide textual inputs for profile updating and reward modeling, we use Qwen3-VL-8B to generate a structured video caption
$c_v=(\mathrm{tag}_v,\mathrm{desc}_v)$ for each video $v$.
Here $\mathrm{tag}_v$ is drawn from a closed-set taxonomy of 25 content categories, and $\mathrm{desc}_v$ is a concise free-text summary of the video content.
Caption generation is performed offline and treated as given in this work.

\subsection{Task Formulation}
\label{sec:task}

\paragraph{Data.}
Let $\mathcal{U}$ and $\mathcal{V}$ denote the sets of users and
short videos, respectively.
Following the engagement criterion used in the production system, an \textbf{effective view} refers to a watch event where the play duration exceeds 7\,seconds or the video is watched to completion.
An observed negative interaction refers to a video that is exposed in the single-column feed but not effectively watched.
For each user $u \in \mathcal{U}$, we collect up to the latest
$N = 500$ effective views before a cutoff time as the behavioral
history
$\mathcal{H}_u = [c_{v_1}, \ldots, c_{v_{|\mathcal{H}_u|}}]$, ordered
chronologically.

\paragraph{Chunked streaming formulation.}
In online recommendation, user behaviors arrive sequentially, and profiles must be updated from the previous state rather than regenerated from scratch.
This streaming setting also makes it impractical to process all $N$ captions in a single LLM call due to context-length constraints.
We therefore partition $\mathcal{H}_u$ into $M$ chronologically
ordered chunks $C_1, \dots, C_M$ (each containing up to $B=50$
captions, with $M=\lceil |\mathcal{H}_u|/B \rceil$) and maintain a
\textbf{user profile} $P_u^t$ that is updated incrementally:
\begin{equation}
  P_u^t \;=\; f_\theta(C_t,\; P_u^{t-1}),
  \qquad t = 1, \dots, M,
  \label{eq:streaming}
\end{equation}
where $P_u^0 = \varnothing$ and $f_\theta$ denotes the learnable
semantic update function parameterized by the LLM policy $\theta$.
The concrete design of $P_u^t$ is presented in
Section~\ref{sec:memory}.

\paragraph{Optimization objective.}
During offline training, RECAP does not interact with the production
recommender or observe online feedback from deployed profiles.
Instead, it uses logged implicit feedback from historical user
behaviors to train a learned proxy reward $R(\cdot)$.
The resulting offline closed-loop optimizes the profile generator
through this proxy reward:

\begin{equation}
  \theta^* = \arg\max_\theta \;
  \mathbb{E}_{u \sim \mathcal{U}}
  \bigl[ R\!\bigl(P_u^M\bigr) \bigr].
  \label{eq:objective}
\end{equation}
The design of $P_u^t$, $R(\cdot)$, and the optimization
procedure are detailed in Section~\ref{sec:method}.

\section{Method}
\label{sec:method}

\subsection{Overview}
\label{sec:overview}

As illustrated in Figure~\ref{fig:overview}, RECAP instantiates the streaming profile optimization objective with three components.
\emph{Structured User Memory} maintains a bounded profile state using LLM-generated semantic diffs and deterministic lifecycle control.
\emph{Recommendation-Aligned Reward} constructs profile-targeted semantic feedback by filtering logged implicit behaviors with an LLM judge and training a dual-tower evaluator over rendered profiles and video captions.
\emph{GRPO Training} optimizes the updater by sampling multiple diffs for each chunk and using evaluator-based rewards with format and diversity constraints.

\subsection{Structured User Memory}
\label{sec:memory}

\paragraph{Profile schema.}
To support controllable long-horizon updates, we represent each profile $P_u^t$ as a fixed-size set of up to $L$ structured interest entries, following recent slot-based user-state designs~\cite{yi2025recgptv2technicalreport}.
Compared with free-form summaries, slot-based entries can be independently confirmed, inserted, or evicted, limiting drift under repeated updates while keeping a predictable token budget.

Each entry contains a semantic pair $(\mathit{topic}_l,\mathit{desc}_l)$ and lifecycle metadata.
The topic is a concise interest label for cross-chunk matching, while the description specifies the user's content preference.
The lifecycle fields, managed only by the state machine and never exposed to the LLM, include \emph{span}, \emph{recency}, and \emph{strength}, summarizing confirmation history, inactivity, and eviction priority.

Lifecycle fields further categorize entries as \textbf{stable} or \textbf{recent}: stable entries capture established long-term interests, while recent entries capture emerging short-term interests.
The LLM only receives ordered $(\mathit{topic},\mathit{desc})$ pairs, while these labels are preserved for profile rendering.

\paragraph{Two-component decomposition.}
We decompose the streaming update $f_\theta$ into an LLM semantic module and a deterministic state machine.
The LLM emits only a minimal \emph{semantic diff}---re-observed existing interests and newly emerged ones---while deterministic rules handle lifecycle bookkeeping and capacity control.
This split is motivated by the observation that LLMs are effective at semantic abstraction but less reliable at maintaining compact and valid numerical state. Directly asking the model to update lifecycle fields can produce inconsistent spans, recencies, or strengths, whereas deterministic rules provide stable bookkeeping while keeping the LLM output focused on semantics.

\paragraph{Semantic module (parameterized by $\theta$).}
Given the current chunk $C_t$ and the
$(\mathit{topic},\;\mathit{desc})$ pairs from $P_u^{t-1}$, the
LLM outputs a structured JSON containing:
\begin{itemize}
  \item \texttt{confirmed\_interests}: existing topics
    re-observed in $C_t$, with optionally refined descriptions;
  \item \texttt{new\_interests}: newly emerged topics discovered
    from $C_t$.
\end{itemize}
Each output interest is associated with \texttt{video\_indices}, i.e., the positions of supporting videos within the input chunk $C_t$.
They ground the LLM output in observed behavior and help the state machine control redundancy, but are not rendered into the profile.
A new entry is inserted only when supported by at least two videos, because insertion consumes limited profile capacity and thus requires stronger evidence than updating an existing entry.

\paragraph{State machine $\mathrm{SM}(\cdot)$ (no learnable parameters).}
The state machine applies three operations to the LLM output and $P_u^{t-1}$:
\textsc{Update} matches confirmed topics to existing entries, refines descriptions when provided, and refreshes or ages their lifecycle fields;
\textsc{Add} filters LLM-proposed new topics, then initializes qualified topics as recent entries; and
\textsc{Remove} recomputes strengths, reassigns stable/recent entries, and evicts low-priority stale entries under the memory budget.
\begin{equation}
  P_u^t \;=\;
  \mathrm{SM}\!\bigl(
    \mathrm{LLM}_\theta(C_t,\, P_u^{t-1}),\;
    P_u^{t-1}
  \bigr).
  \label{eq:update}
\end{equation}

\paragraph{Profile-to-text assembly.}
For downstream use, $P_u^t$ is rendered as $s(P_u^t)$, listing stable entries under \emph{Core Interests} and recent entries under \emph{Recent Interests}, each formatted as \texttt{topic: description}.
The rendered text is used by the dual-tower evaluator and downstream recommendation models.

\subsection{Recommendation-Aligned Reward}
\label{sec:reward}
To optimize the profile generator offline, we construct a profile-targeted proxy reward from historical behaviors.
Rather than using production-ranker outputs, which mix profile, ID, context, popularity, and ranking signals, we train a dual-tower evaluator over the rendered profile $s(P_u)$ and video caption $c_v$ as a controlled profile-item utility signal.

\subsubsection{Dual-tower semantic evaluator.}
The evaluator encodes $s(P_u)$ and $c_v$ with a user tower and an
item tower, respectively, and projects them into a shared embedding
space.
It contains a ranking head, whose embeddings
$\mathbf{e}_u^{\mathrm{r}}, \mathbf{e}_v^{\mathrm{r}}$ are used
for effective-view prediction, and a contrastive head, whose
normalized embeddings
$\mathbf{e}_u^{\mathrm{c}}, \mathbf{e}_v^{\mathrm{c}}$ are used
for semantic separation.
The evaluator is trained with
\begin{equation}
  \mathcal{L}_{\mathrm{eval}}
    =
    w_{\mathrm{bce}}\,\mathcal{L}_{\mathrm{bce}}
    +
    w_{\mathrm{nce}}\,\mathcal{L}_{\mathrm{nce}},
  \label{eq:eval_loss}
\end{equation}
where $\mathcal{L}_{\mathrm{bce}}$ predicts effective views and
$\mathcal{L}_{\mathrm{nce}}$ applies a contrastive loss on positive
interactions with distributed in-batch negatives~\cite{oord2018representation,radford2021learning}.
Together, the two losses support both implicit-feedback discrimination and retrieval-style semantic comparison.

\subsubsection{High-confidence feedback construction.}
Because watch and skip labels are implicit behavior signals rather than explicit semantic preferences, we first estimate their consistency with the user's recent interests.
We construct cleaned behavioral data via LLM-judge consistency filtering, as illustrated in Figure~\ref{fig:overview}.

For each user, the LLM judge observes recent effective-view history before the candidate interaction window and chooses, from a randomized positive--negative video pair, which item the user is more likely to effectively watch.
Aggregating multiple pairwise comparisons gives each item a label-consistency score \(q_{\mathrm{cons}}\), indicating how often the judge's context-conditioned preference agrees with the observed feedback label.
To avoid using unreliable negatives as references for positives, we first identify high-confidence negatives and then re-estimate positive consistency only against these confirmed negatives.
Samples with \(q_{\mathrm{cons}}>\tau\) are retained as cleaned behavioral data for training cleaned evaluators and computing cleaned-feedback rewards.

\subsubsection{Recommendation-aligned reward.}
For a candidate intermediate profile $\tilde{P}_{u,i}^t$ obtained after
processing chunk $C_t$, we score its semantic compatibility
with a cleaned reward set
\(\mathcal{V}_u = \mathcal{V}_u^+ \cup \mathcal{V}_u^-\), where
\(\mathcal{V}_u^+\) and \(\mathcal{V}_u^-\) denote positive and negative videos retained after label-consistency filtering.
The recommendation-aligned reward is defined as
\begin{equation}
  R_{\mathrm{rec}}(\tilde{P}_{u,i}^t)
    =
    \exp\!\Bigl(
      -\mathcal{L}_{\mathrm{bce}}
      \bigl(s(\tilde{P}_{u,i}^t), \mathcal{V}_u\bigr)
    \Bigr)
    +
    w_{\mathrm{rec}} \cdot
    \Delta\!\cos
    \bigl(s(\tilde{P}_{u,i}^t), \mathcal{V}_u\bigr),
  \label{eq:r_rec}
\end{equation}
where \(w_{\mathrm{rec}}\) controls the weight of the cosine-margin
reward term. The first term measures how well the profile predicts the
corresponding feedback labels, and the second term measures the
semantic separation between positive and negative video items:
\begin{equation}
  \Delta\!\cos
  \bigl(s(\tilde{P}_{u,i}^t), \mathcal{V}_u\bigr)
    =
    \frac{1}{|\mathcal{V}_u^+|}
    \sum_{v \in \mathcal{V}_u^+}
      \cos(\mathbf{e}_{\tilde{P}}^{\mathrm{c}},
           \mathbf{e}_v^{\mathrm{c}})
    -
    \frac{1}{|\mathcal{V}_u^-|}
    \sum_{v \in \mathcal{V}_u^-}
      \cos(\mathbf{e}_{\tilde{P}}^{\mathrm{c}},
           \mathbf{e}_v^{\mathrm{c}}).
  \label{eq:cos_margin}
\end{equation}
The evaluator is trained once and kept frozen; its score provides the recommendation-aligned reward used in GRPO, before adding output-control penalties.
\subsection{GRPO Training}
\label{sec:grpo}

\paragraph{SFT initialization.}
Before policy optimization, we warm-start the profile generator with SFT demonstrations generated by a stronger teacher model under the same structured output schema.
The resulting SFT checkpoint initializes $\pi_\theta$ and serves as the frozen reference policy $\pi_{\mathrm{ref}}$.

\paragraph{Parallel chunk-level rollouts.}
Directly optimizing the streaming process requires sequential
rollouts over all $M$ chunks for each user.
To reduce this cost, we pre-compute prefix profiles with the SFT
model and optimize each chunk-level update independently.
For user $u$ and chunk $t$, we construct a training instance
$\langle C_t, \bar{P}_u^{t-1} \rangle$, where
$\bar{P}_u^{t-1}$ is the profile obtained from preceding chunks by
the SFT model.
The current policy is responsible only for generating the semantic
diff for chunk $C_t$, enabling parallel rollouts across users and
chunks.

\paragraph{Reward composition.}
For each chunk-level input $\langle C_t, \bar{P}_u^{t-1} \rangle$, the
policy samples $K$ candidate semantic diffs
$\{O_{u,1}^t,\ldots,O_{u,K}^t\}$.
Each diff is applied to the deterministic state machine to obtain a
candidate updated profile:
\begin{equation}
  \tilde{P}_{u,i}^t
  =
  \mathrm{SM}(O_{u,i}^t,\bar{P}_u^{t-1}).
  \label{eq:rollout_profile}
\end{equation}
The rollout reward combines the recommendation-aligned reward
with output-control penalties:
\begin{equation}
  r_{u,i}^t
  =
  R_{\mathrm{rec}}(\tilde{P}_{u,i}^t)
  -
  \lambda_{\mathrm{fmt}} P_{\mathrm{fmt}}(O_{u,i}^t)
  -
  \lambda_{\mathrm{div}} P_{\mathrm{div}}(\tilde{P}_{u,i}^t).
  \label{eq:total_reward}
\end{equation}
$P_{\mathrm{fmt}}$ penalizes invalid JSON, mismatched confirmed-interest references, and over-length descriptions, so that the semantic diff remains executable by the state machine and concise enough for bounded profile updates.
$P_{\mathrm{div}}$ penalizes pairs of interest entries with highly overlapping supporting \texttt{video\_indices}, treating evidence overlap as a soft redundancy signal.
This is designed for locally concentrated behavior chunks, where many watched videos may reflect the same dominant content type and should not consume multiple profile slots with near-duplicate preference descriptions.

\paragraph{Group-relative policy optimization.}
Within each rollout group, rewards are normalized into
group-relative advantages:
\begin{equation}
  \hat{A}_{u,i}^t
  =
  \frac{
    r_{u,i}^t - \mathrm{mean}_{j}(r_{u,j}^t)
  }{
    \mathrm{std}_{j}(r_{u,j}^t) + \epsilon_{\mathrm{std}}
  }.
  \label{eq:advantage}
\end{equation}
Following GRPO~\cite{shao2024deepseekmath}, we update the policy
with a clipped surrogate objective and a KL penalty to the frozen
reference policy:
\begin{equation}
\begin{aligned}
  \mathcal{L}_{\mathrm{GRPO}}(\theta)
  ={}&
  - \mathbb{E}_{u,t,i,j}
  \left[
    \ell_{u,i,j}^t
  \right]
  +
  \beta D_{\mathrm{KL}}
  \bigl(\pi_\theta \,\|\, \pi_{\mathrm{ref}}\bigr), \\
  \ell_{u,i,j}^t
  ={}&
  \min\!\Bigl(
    \rho_{u,i,j}^t \hat{A}_{u,i}^t,\;
    \mathrm{clip}(\rho_{u,i,j}^t,1-\epsilon_{\mathrm{clip}},1+\epsilon_{\mathrm{clip}})
    \hat{A}_{u,i}^t
  \Bigr).
\end{aligned}
\label{eq:grpo}
\end{equation}
Here $j$ indexes output tokens, and
$\rho_{u,i,j}^t$ is the probability ratio between the current
policy and the rollout policy.

\section{Experiments}
\label{sec:experiments}

Our experiments are designed to answer the following research questions:
\begin{itemize}
    \item \textbf{RQ1:} Does feedback-driven optimization improve the utility of streaming structured semantic profiles in offline dual-tower evaluation?
    \item \textbf{RQ2:} How do behavioral-data cleaning and reward design affect profile optimization?
    \item \textbf{RQ3:} Are memory-update decomposition and profile capacity important for stable and practical profile generation?
    \item \textbf{RQ4:} How do different training signals change the behavior and semantics of generated profiles?
\end{itemize}

\begin{table}[t]
\centering
\caption{Statistics of the training and target-window data.}
\label{tab:data_stats}
\begin{tabular}{@{}llrrrr@{}}
\toprule
User Set & Data & Users & Items & \# Train  & \# Target  \\
\midrule
Policy   & Raw   & 10K & 1,639,176 & 2,000,000 & 5,558,911 \\
Policy   & Cleaned & 10K & 1,288,972 & 1,298,655 & 3,609,485 \\
Held-out & Raw   & 10K & 1,643,388 & 2,000,000 & 5,644,058 \\
Held-out & Cleaned & 10K & 1,290,175 & 1,295,564 & 3,659,622 \\
\bottomrule
\end{tabular}
\end{table}

\subsection{Experimental Setup}
\label{sec:exp_setup}

\paragraph{Dataset construction.}
We evaluate on an internal Kuaishou short-video recommendation dataset with two disjoint 10K-user sets: a policy set for reward computation and GRPO optimization, and a held-out set used only for final profile utility evaluation.
For each user, interactions are temporally divided into a profile window, a training window, and a later target window, which serves as the reward window on the policy set and the evaluation window on the held-out set.
The profile window provides up to the latest 500 historical effective views for streaming profile construction, with an average length of 463.
The training and target windows provide user-item pairs for dual-tower training and downstream reward/evaluation scoring, respectively; training pairs are capped at the latest 200 interactions per user, yielding up to 2M pairs for each 10K-user set.
For both user sets, we construct raw behavioral data from observed implicit interactions and cleaned behavioral data using the LLM-judge consistency filtering process.
Table~\ref{tab:data_stats} summarizes the resulting data statistics.

\paragraph{Training and evaluation protocol.}
We use the same dual-tower architecture for reward construction and final offline evaluation, but instantiate and train separate evaluators on disjoint user sets.

For policy optimization, the policy set is used to train reward evaluators and compute GRPO rewards on its reward window.
For each reward setting, the reward evaluator is trained on the training window and then frozen to score candidate profiles on the reward window.
GRPO-clean and GRPO-raw use reward evaluators trained with cleaned and raw behavioral data, respectively.
Since the evaluator takes generated user profiles as user-side inputs, its training profiles are generated by the corresponding initialization model, so that the evaluator input distribution matches the starting policy distribution.

For profile utility evaluation, all profile generators are evaluated on the held-out user set.
Each generator first performs streaming inference on the profile window of these users to produce final profiles, which are then used as user-side semantic inputs for training and evaluating the offline dual-tower evaluator.
For each data setting, the evaluator is trained and evaluated with the corresponding behavioral data: cleaned evaluation uses cleaned training and evaluation pairs, while raw evaluation uses raw training and evaluation pairs.
Thus, all variants share the same held-out users, temporal split, captions, evaluator architecture, hyperparameters, and metrics; only the profile generator changes.

\paragraph{Compared variants and metrics.}
We compare five controlled variants under the same structured profile framework.
Base denotes the original profile generator.
GRPO-raw and GRPO-clean further optimize the Base generator with raw-feedback and cleaned-feedback rewards, respectively.
SFT is trained with teacher-generated demonstrations.
RECAP (SFT+GRPO) is the full model, which initializes the policy with SFT and further optimizes it with cleaned-feedback rewards.

We report uAUC and Recall@\(2000\) for offline dual-tower evaluation on both cleaned and raw evaluation data.
We use Recall to denote Recall@\(2000\) hereafter.
For profile behavior analysis, we further compute programmatic diagnostics from the structured model outputs, including historical coverage, duplicate assignment rate, update rate, and empty-interest count.

\paragraph{Implementation details.}
The profile generator is initialized from Qwen3-4B-Instruct-2507~\cite{yang2025qwen3technicalreport}.
SFT uses 8K teacher-generated demonstrations from Claude Sonnet 4.5.
GRPO is trained for 80 steps with batch size 64, learning rate \(2.5\times 10^{-5}\), LoRA rank 32, KL coefficient 0.01, and \(K=8\) sampled outputs per instance.
In Eq.~\eqref{eq:r_rec}, $w_{\mathrm{rec}}$ weights the cosine-margin reward term and is set to $1$ in all experiments.
The dual-tower evaluator uses Qwen3-Embedding-0.6B to encode both user profiles and video captions, followed by task-specific projection heads; we set \(w_{\mathrm{bce}}=1\) and \(w_{\mathrm{nce}}=0.1\) for evaluator training.
The LLM judge used for consistency filtering is Qwen3-30B-A3B-Instruct-2507, and its first valid A/B token is parsed as the pairwise preference.
We set the consistency threshold $\tau$ to $0.45$.

\subsection{Overall Performance}
\label{sec:overall}

\begin{table}[t]
\centering
\small
\caption{Overall performance in offline dual-tower evaluation on the held-out user set.}
\label{tab:overall}
\begin{tabular}{@{}lcccc@{}}
\toprule
Method 
& \multicolumn{2}{c}{Cleaned Eval.} 
& \multicolumn{2}{c}{Raw Eval.} \\
\cmidrule(lr){2-3} \cmidrule(lr){4-5}
& uAUC & Recall & uAUC & Recall \\
\midrule
Base       & 0.7519 & 0.0122 & 0.5871 & 0.0088  \\
GRPO-raw   & 0.7539 & 0.0119 & 0.5874 & 0.0087  \\
GRPO-clean & 0.7572 & 0.0121 & 0.5883 & 0.0087  \\
\midrule
SFT        & 0.7564 & 0.0127 & 0.5866 & \textbf{0.0091}  \\
RECAP (SFT+GRPO)  & \textbf{0.7603} & \textbf{0.0128} & \textbf{0.5886} & 0.0089  \\
\bottomrule
\end{tabular}
\end{table}

\begin{figure*}[t]
    \centering
    \includegraphics[width=0.98\textwidth]{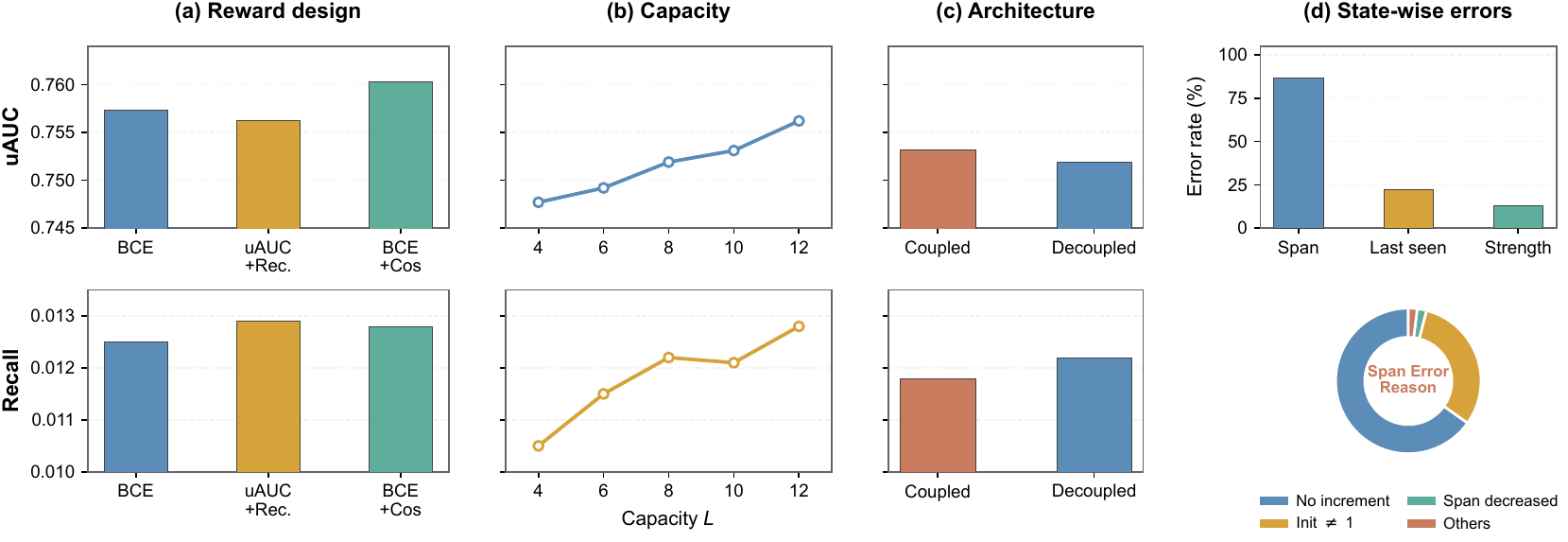}
    \Description{Four ablation panels compare reward formulations, profile capacities, coupled and decoupled update architectures, and state-field errors using uAUC, Recall, error rates, and span-error reasons.}
    \caption{Ablation studies of RECAP. 
    (a) compares different reward formulations. 
    (b) evaluates different profile capacities \(L\). 
    (c) studies the effect of decoupling semantic updates from state management. 
    (d) analyzes state-field errors in the coupled design, including field-wise error rates and span-error reasons.}
    \label{fig:ablation}
\end{figure*}

Table~\ref{tab:overall} reports the offline dual-tower evaluation results of different profile generators on the held-out user set.
We report results on both cleaned and raw evaluation data: cleaned evaluation provides a higher-confidence view of semantic profile--video alignment, while raw evaluation retains the original implicit behavioral distribution and avoids evaluating only on the cleaned subset.
On cleaned evaluation, RECAP (SFT+GRPO) achieves the best performance on both uAUC and Recall.
Compared with Base, it improves cleaned-evaluation uAUC from 0.7519 to 0.7603 and Recall from 0.0122 to 0.0128.

Compared with GRPO-raw, GRPO-clean achieves higher cleaned-evaluation performance and also improves raw-evaluation uAUC from the same Base generator.
This suggests that filtering ambiguous implicit feedback provides a cleaner reward signal without merely overfitting to the cleaned subset.
At the dual-tower evaluation level, these results show that SFT initialization with cleaned-feedback GRPO gives the strongest profile--video alignment; Section~\ref{sec:behavior} further analyzes how this improvement appears in the generated profiles.

\subsection{Ablation Studies}

\label{sec:ablation}
\begin{figure}[t]
\centering
\footnotesize
\setlength{\tabcolsep}{3pt}
\begin{tabular}{p{0.22\linewidth}p{0.72\linewidth}}
\toprule
Method & Generated interest for the same relationship-guidance behavior cluster \\
\midrule
GRPO-raw &
\textbf{Relationship Tarot Readings.}
Enjoys tarot sessions with direct address, dramatic visuals, and spiritual or mystical framing around jealousy, reconciliation, and past relationships. \\

GRPO-clean &
\textbf{Romantic Relationship Predictions.}
Engages with tarot-based readings about love and relationship dynamics, especially emotionally charged predictions about past relationships, conflicts, and potential reconnections. \\

SFT+GRPO &
\textbf{Relationship Conflict Resolution Stories.}
Aggregates tarot, horoscope, and relationship-advice content into emotional guidance about healing, proactive opportunity creation, independence, and self-worth. \\
\bottomrule
\end{tabular}
\Description{A three-row qualitative comparison of the profile descriptions generated by GRPO-raw, GRPO-clean, and SFT+GRPO for the same relationship-guidance behavior cluster.}
\caption{Qualitative comparison of generated profile descriptions for the same relationship-guidance behavior cluster from one sampled user.}
\label{fig:qual_case}
\end{figure}

Figure~\ref{fig:ablation} studies the main design choices in RECAP, including reward formulation, profile capacity, memory-update decomposition, and state-field reliability.

\subsubsection{Reward design.}
Figure~\ref{fig:ablation}(a) compares different reward formulations for GRPO.
The combined BCE and cosine-margin reward achieves the strongest overall performance in offline dual-tower evaluation.
This suggests that effective profile optimization requires both label-level discrimination and semantic separation between positive and negative items.
Using only BCE weakens the retrieval-oriented semantic signal, while directly optimizing coarse evaluation metrics provides a less stable reward for chunk-level policy updates.

\subsubsection{Profile capacity.}
\label{sec:capacity}
Figure~\ref{fig:ablation}(b) studies the maximum number of interest entries \(L\).
Larger capacities can represent more diverse interests and improve offline dual-tower performance, but increase the cost of streaming generation, GRPO rollouts, profile encoding, inspection, and maintenance.
Under our training and serving budget, \(L=8\) balances interest diversity with profile compactness and is used by default.

\begin{figure*}[t]
    \centering
    \includegraphics[width=0.95\textwidth]{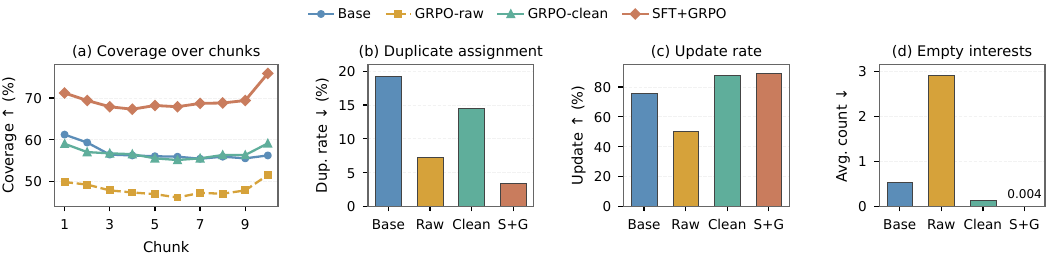}
    \Description{Four panels compare Base, GRPO-raw, GRPO-clean, and SFT+GRPO using coverage over historical chunks, duplicate assignment rate, update rate, and average empty-interest count.}
    \caption{Profile behavior diagnostics under different training signals.
    (a) Coverage is the LLM-judged fraction of videos in each historical chunk covered by the final generated profile.
    (b) Duplicate assignment rate measures how often the same historical video is assigned to multiple interest entries.
    (c) Update rate measures how often existing interest descriptions are refined.
    (d) Empty interests are generated entries with no supporting videos.
    In the bar plots, Raw, Clean, and S+G abbreviate GRPO-raw, GRPO-clean, and SFT+GRPO, respectively.}
    \label{fig:profile_behavior}
\end{figure*}

\begin{figure}[t]
    \centering
    \includegraphics[width=\linewidth]{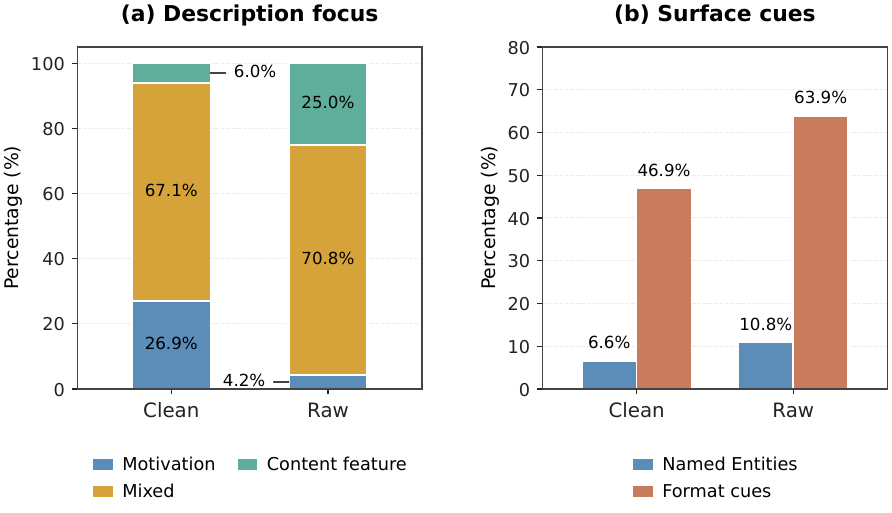}
    \Description{Two bar charts compare GRPO-clean and GRPO-raw profiles by description focus and by the frequency of named entities and video-format cues.}
    \caption{LLM-as-judge annotation of GRPO-clean and GRPO-raw profiles.
    (a) Description focus classifies each generated interest as motivation-focused, content-feature-focused, or mixed.
    (b) Surface cues report how often generated interests explicitly mention named entities or video-format cues.}
    \label{fig:semantic_judge}
\end{figure}

\subsubsection{Memory-update decomposition.}
Figure~\ref{fig:ablation}(c) compares RECAP with a coupled full-state updater, where the LLM directly generates the complete updated profile, including both interest semantics and lifecycle fields.
The coupled updater is a competitive compact semantic-summary variant and achieves comparable offline dual-tower performance to the decoupled design.
However, Figure~\ref{fig:ablation}(d) shows that directly generating lifecycle fields causes frequent state-transition errors, especially for cross-chunk span updates, making the resulting profile difficult to trace over time.
This supports the decoupled design: the LLM handles semantic updates, while deterministic rules manage recency, span, strength, and capacity control.

\subsection{Profile Behavior Analysis}
\label{sec:behavior}

\subsubsection{Diagnostic Protocol.}
We complement offline dual-tower evaluation with profile-level behavior diagnostics.
We use gpt-oss-120b~\cite{openai2025gptoss120b} for LLM-based semantic judgments, including historical coverage and semantic annotation.
For coverage, the judge assigns each historical video to a generated interest or to an \texttt{unknown} category, and coverage is the fraction assigned to non-\texttt{unknown} interests.
For semantic annotation, we first ask the LLM judge to compare sampled GRPO-clean/GRPO-raw profile pairs in an open-ended manner and summarize recurring differences; the two dominant dimensions, description focus and reliance on surface textual cues, are then used as annotation criteria.
The remaining diagnostics are computed programmatically from raw structured outputs before support-based filtering: duplicate assignment from overlapping \texttt{video\_indices}, update rate from refined confirmed interests, and empty interests from proposed entries with no supporting videos.
Coverage is computed on 500 users, while semantic annotation is performed on 500 sampled chunks for content diversity.
All diagnostics are used only for analysis and are not used as training rewards.

\subsubsection{Update Behavior under Raw and Cleaned Rewards.}
Figure~\ref{fig:profile_behavior}(c) and Figure~\ref{fig:profile_behavior}(d) reveal a failure mode under raw-feedback rewards.
GRPO-raw updates existing interest descriptions much less frequently than GRPO-clean, while producing many more empty interests, i.e., proposed entries with no supporting videos before filtering.
This suggests that raw-feedback rewards fail to provide a stable optimization direction for profile editing.
One interpretation is that, under an uncertain reward signal, leaving existing interests unchanged or proposing unsupported entries can become low-risk actions: the former avoids harmful rewrites, while the latter is removed before the state machine update.
In contrast, cleaned-feedback rewards provide a clearer optimization direction, so the policy more often makes grounded edits, either by refining existing interests or by proposing new interests supported by the current chunk.

\subsubsection{Semantic Focus under Cleaned-Feedback Rewards.}
Beyond update behavior, we examine the semantics captured by generated profiles.
Figure~\ref{fig:semantic_judge}(a) shows that GRPO-clean produces more motivation-focused descriptions of why users watch, whereas GRPO-raw produces more content-feature-focused descriptions of what appears in the videos.
Consistently, Figure~\ref{fig:semantic_judge}(b) shows that raw-feedback profiles mention named entities and video-format cues more frequently, suggesting that raw-feedback rewards encourage surface-feature matching while cleaned-feedback rewards enable deeper semantic abstraction.
This also helps interpret duplicate assignment in Figure~\ref{fig:profile_behavior}(b): motivation-level descriptions may overlap when related user intents explain similar videos, whereas surface-feature descriptions may appear more separated through concrete entities or format cues without necessarily modeling user interests better.

\subsubsection{Complementary Effects of SFT and GRPO.}
These diagnostics provide a profile-level explanation for the dual-tower gains in Table~\ref{tab:overall}: SFT and GRPO play complementary roles in profile generation.
Figure~\ref{fig:profile_behavior}(a--b) shows that SFT+GRPO achieves the highest historical coverage and the lowest duplicate assignment rate, indicating that supervised demonstrations improve both behavioral coverage and separation among interests.
In contrast, GRPO shapes profile editing and abstraction under recommendation feedback: the GRPO-clean/GRPO-raw comparison shows that reward quality determines whether the model performs grounded refinement or resorts to low-impact, surface-level updates.
Figure~\ref{fig:qual_case} provides a consistent qualitative example: GRPO-raw emphasizes surface format, GRPO-clean abstracts relationship-oriented viewing motivation, and SFT+GRPO further aggregates related formats into a broader interest.
Overall, SFT improves profile organization, while GRPO with cleaned-feedback rewards improves how the profile is updated and abstracted.

\subsection{Online A/B Test}
\label{sec:online_ab}

We conduct a seven-day online A/B test on Kuaishou's overseas short-video recommendation service, allocating 10\% of the traffic to the treatment group and 20\% to the control group.
For treatment users, RECAP-generated user memory is rendered as text, encoded using Qwen3-Embedding-0.6B, and provided to the production U2I retrieval model as an additional feature; the control group uses the original model without this feature.
Compared with the control group, RECAP yields a statistically significant relative improvement of 0.139\% in average application usage time per user.

\section{Conclusion}
\label{sec:conclusion}

Streaming structured semantic profiles provide an interpretable state for evolving user interests, but optimizing a bounded profile updater from industrial implicit feedback remains challenging.
We propose RECAP, an offline closed-loop framework that maintains bounded structured user memory, constructs profile-targeted semantic rewards from label-consistency-filtered behavior pairs, and optimizes LLM-based profile updates with GRPO.
Experiments on real-world Kuaishou short-video data show that RECAP improves profile utility in offline evaluation.
Further analyses verify that cleaned-feedback rewards produce more grounded and semantically abstract profile refinements, while decoupled lifecycle management improves the reliability of streaming state maintenance.
An online A/B test further confirms the practical effectiveness of RECAP in real-world short-video recommendation.

\begin{acks}
This work was supported by the National Natural Science Foundation of China (62402470), the Fundamental Research Funds for the Central Universities of China (WK2100000053), and the Anhui Provincial Natural Science Foundation (2408085QF189).
This work was also supported by the Open Innovation Program for jointly trained master's-to-doctoral students at the Institute of Computing Technology, Chinese Academy of Sciences, and by the advanced computing resources provided by the Supercomputing Center of USTC.
\end{acks}




\bibliographystyle{ACM-Reference-Format}
\bibliography{reference}

\end{document}